\documentclass{epl}

\newcommand{\Prob}{\mbox{Prob}}
\newcommand{\sgn}{\mbox{sgn}}
\renewcommand{\vec}[1]{\mbox{\boldmath $#1$}}

\title{Low-dimensional chaos induced by frustration in a non-monotonic
system}
\shorttitle{Low-dimensional chaos induced by frustration}
\author{M. Kawamura\inst{1} \and R. Tokunaga\inst{2} \and
M. Okada\inst{3}}
\institute{
  \inst{1} Faculty of Science, Yamaguchi University \\
  Yoshida 1677-1, Yamaguchi, 753-8512 Japan \\
  \inst{2} Institute of Information Sciences and Electronics,
  University of Tsukuba \\
  Tennodai 1-1-1, Tsukuba, 305-8573 Japan \\
  \inst{3} Brain Science Institute, RIKEN \\ Wako-shi, 351-0198 Japan
}


\begin{document}
\maketitle

\begin{abstract}
 We report a novel mechanism for the occurrence of chaos at the
 macroscopic level induced by the frustration of interaction, namely
 frustration-induced chaos, in a non-monotonic sequential associative
 memory model. We succeed in deriving exact macroscopic dynamical
 equations from the microscopic dynamics in the case of the
 thermodynamic limit and prove that two order parameters dominate this
 large-degree-of-freedom system. Two-parameter bifurcation diagrams are
 obtained from the order-parameter equations. Then we analytically show
 that the chaos is low-dimensional at the macroscopic level when the
 system has some degree of frustration, but that the chaos definitely
 does not occur without the frustration.
\end{abstract}

The occurrence of chaotic behaviors at the macroscopic level has been
extensively investigated, and thus its mechanisms can be classified into
several classes according to their typical systems. Examples of the
first class are the globally coupled map (GCM) and chaotic neural
networks whose processing units are themselves chaotic, and the
macroscopic behaviors are also chaotic \cite{Kaneko1989, Shibata1998,
Aihara1990, Adachi1997}.  Since the processing units are chaotic, the
chaos at the macroscopic level seems to be high-dimensional. However,
large-scale computer simulations show that the dimension of the chaos in
the GCM can be lower than the degree of freedom \cite{Shibata1998}. In
the other class of mechanisms, the processing units are not chaotic
themselves, and thus resultant macroscopic chaotic behaviors would seem
to be induced by other properties. A typical example of this class is
that of neural networks. Since their processing units are simple binary
units, their chaotic behaviors should be induced by additional
mechanisms, that is, synaptic pruning, synaptic delay, thermal noise,
sparse connections and/or so on
\cite{FukaiShiino1990,Nara1992,vanVreeswijk1996}. Frustration is worth
notice as inducing chaos. For continuous systems, chaotic behaviors can
be analyzed in some small networks with frustration of interaction at
the microscopic level
\cite{Bersini1995,Bersini1998,Bersini2002} and random neural
networks at the macroscopic level \cite{Sompolinsky1988}.  Here, we
report a novel mechanism of the occurrence of chaos at the macroscopic
level induced by the frustration of interaction, namely {\itshape
frustration-induced chaos}, for discrete system.

We discuss a non-monotonic sequential associative memory model. The
sequential associative memory model is a neural network, in which the
sequence of patterns is embedded as an attractor through the Hebbian
(correlation) learning
\cite{Amari1988,During1998,Katayama2001,Kawamura2002}. When the number
of the patterns, $p$, is on the order ${\cal O}(N)$, where $N$ is the
number of the processing units, the model has frustrated interactions
\cite{Rieger1998}. Previously, the properties at stationary states had
been exactly analyzed by the path-integral method
\cite{During1998,Katayama2001} because the theoretical treatment of
transient dynamics was difficult. However, we recently succeeded in
rigorously analyzing the transient dynamics of the
model~\cite{Kawamura2002}.

It has been reported that the non-monotonicity of processing units (a
larger absolute value of local field tends to make their state opposite
to the local field) gives the systems some superior properties, e.g., an
enhancement of the storage capacity, few spurious state, and a super
retrieval phase \cite{Morita1993,ShiinoFukai1993,Okada1996}. An
interesting one is chaotic behavior: Results of numerical simulations
showed that the systems with non-monotonic units have chaotic
behaviors. In particular, chaotic behavior was observed when the
retrieval process failed. Dynamical theories are indispensable to
analyzing these chaotic behaviors. Only approximated theories, e.g.,
Gaussian approximation
\cite{Amari1988,Okada1996,AmariMaginu1988,Okada1995,
Kawamura1999,NishimoriOpris1993} or steady-state approximation
\cite{During1998, Katayama2001}, have been employed to investigate the
occurrence of these chaotic behaviors theoretically
\cite{NishimoriOpris1993, Katayama2001}, since theoretical treatment was
difficult as mentioned above. Consequently, the occurrence of these
chaotic behaviors has never been analyzed rigorously.

In this Letter, we extend our exact dynamical theory \cite{Kawamura2002}
for the sequential associative memory model to a non-monotonic one to
discuss the occurrence of chaos. We have succeeded in obtaining an exact
description of its macroscopic dynamics and proved that two order
parameters dominate this model. We obtained two-parameter bifurcation
diagrams from the order-parameter equations and analytically
demonstrated that the chaos of the present system is low-dimensional at
the macroscopic level. We also analytically show that the chaotic
behavior is observed when the retrieval process fails, and prove that
the chaotic behaviors in this system occur only when it has some degree
of frustration. This means that this macroscopic chaos is induced by the
frustration, that is, it is {\itshape frustration-induced chaos}.

Let us consider a sequential associative memory model that consists of
$N$ units or neurons. The state of the units takes $\sigma_i(t)=\pm1$
and updates the state synchronously with the following probability:
\begin{equation}
 \Prob\left[\sigma_i(t+1)\vert h_i(t)\right]
  =\frac12\left[1+\sigma_i(t+1)F\left(h_i(t)\right) \right] ,
  \label{eqn:dynamics}  
\end{equation}
\begin{equation}
 h_i(t)=\sum_{j=1}^NJ_{ij}\sigma_j(t)+I_i(t) ,
  \label{eqn:hi} 
\end{equation}
where $J_{ij}$ is the coupling, $I_i(t)$ is the threshold or external
input, and $h_i(t)$ is the local field. The function $F\left(h\right)$
is a non-monotonic function given by
\begin{equation}
 F\left(h\right)=\tanh\beta h-\tanh\beta\left(h-\theta\right)
  -\tanh\beta\left(h+\theta\right) ,
\end{equation}
where $\beta$ is the inverse temperature $\beta=1/T$, and $\theta$ is
the non-monotonicity. When $T\to0$, update rule of the model becomes
deterministic, 
\begin{equation}
 \sigma_i(t+1)=\sgn\left(h_i(t)\right)-\sgn\left(h_i(t)-\theta\right)-\sgn\left(h_i(t)+\theta\right) .
\end{equation}
When the absolute value of the local field is larger then $\theta$, the
state has opposite sign to the local field, i.e., 
$\sigma_i(t+1)=-\sgn\left(h_i(t)\right)$. The coupling $J_{ij}$ stores
$p$ random patterns $\vec{\xi}^{\mu}=(\xi^{\mu}_1,\cdots,\xi^{\mu}_N)^T$
so as to retrieve the patterns as
$\vec{\xi}^0\to\vec{\xi}^1\to\cdots\vec{\xi}^{p-1}\to\vec{\xi}^0$
sequentially. It is given by
\begin{equation}
 J_{ij}=\frac1N\sum_{\mu=0}^{p-1}\xi^{\mu+1}_i\xi^{\mu}_j ,
  \label{eqn:Jij}
\end{equation}
where $\vec{\xi}^{p}=\vec{\xi}^0$. The number of stored patterns $p$ is
given by $p=\alpha N$, where $\alpha$ is called the {\itshape loading
rate}. Each component of the patterns is assumed to be an independent
random variable that takes a value of either $+1$ or $-1$ according to
the following probability,
\begin{equation}
 \Prob\left[\xi_i^{\mu}=\pm1\right]=\frac{1}{2} .
\end{equation}
We determine the initial state $\vec{\sigma}(0)$ according to the
following probability distribution,
\begin{equation}
 \Prob[\sigma_i(0)=\pm1] = \frac{1\pm m(0) \xi^0_i}{2}. 
\end{equation}
The overlap being the direction cosine between $\vec{\sigma}(0)$ and 
$\vec{\xi}^0$ converges to $m(0)$ when $N\to\infty$.

In order to discuss the transient dynamics, we introduce macroscopic
state equations by the path-integral method
\cite{During1998,Katayama2001,Kawamura2002}. The generating function
$Z[\vec{\psi}]$ is defined as
\begin{eqnarray}
 Z[\vec{\psi}] &=& 
 \sum_{\vec{\sigma}(0),\cdots,\vec{\sigma}(t)} 
 p\left[\vec{\sigma}(0),\vec{\sigma}(1),\cdots,\vec{\sigma}(t)\right]
  \exp\left(-i\sum_{s<t}\vec{\sigma}(s)\cdot\vec{\psi}(s)\right) ,
 \label{eqn:Z0} 
\end{eqnarray}
where $\vec{\psi}=\left(\vec{\psi}(0),\cdots,\vec{\psi}(t-1)\right)$.
The state $\vec{\sigma}(s)=(\sigma_1(s),\cdots,\sigma_N(s))^T$ denotes
state of the spins at time $s$, and the path probability
$p\left[\vec{\sigma}(0),\vec{\sigma}(1),\cdots,\vec{\sigma}(t)\right]$
denotes the probability of taking the path from initial state
$\vec{\sigma}(0)$ to state $\vec{\sigma}(t)$ at time $t$ through
$\vec{\sigma}(1),\vec{\sigma}(2),\cdots,\vec{\sigma}(t-1)$. Since the
dynamics (\ref{eqn:dynamics}) is a Markov chain, the path probability
is given by
\begin{equation}
 p\left[\vec{\sigma}(0),\vec{\sigma}(1),\cdots,\vec{\sigma}(t)\right]
  =p\left[\vec{\sigma}(0)\right]\prod_{s<t}
  \prod_{i}\frac{1}{2}\left[1+\sigma_i(s+1)F\left(h_i(s)\right)\right] .
\end{equation}

The generating function $Z[\vec{\psi}]$ involves the following order
parameters:
\begin{eqnarray}
 m(s) &=& i\lim_{\vec{\psi}\to0}\frac{1}{N}\sum_{i=1}^N \xi_i^s
  \frac{\partial Z[\vec{\psi}]}{\partial \psi_i(s)} 
  \label{eqn:def_m} , \\
 G(s,s') &=& i\lim_{\vec{\psi}\to0}\frac{1}{N}\sum_{i=1}^N
  \frac{\partial^2 Z[\vec{\psi}]}{\partial\psi_i(s)\partial I_i(s')} , \\ 
 C(s,s') &=& -\lim_{\vec{\psi}\to0}\frac{1}{N}\sum_{i=1}^N
  \frac{\partial^2 Z[\vec{\psi}]}{\partial\psi_i(s)\partial\psi_i(s')} 
  \label{eqn:def_C} .
\end{eqnarray}
The order parameter $m(s)$ corresponds to the overlap, which represents
the direction cosine between the state $\vec{\sigma}(s)$ and the
retrieval pattern $\vec{\xi}^s$ at time $s$. $G(s,s')$ and $C(s,s')$ are
the response function and correlation function between time $s$ and
$s'$, respectively. Therefore, the problem of discussing the macroscopic
transient dynamics eventually results in the problem of evaluating the
generating function. We consider the case of thermodynamic limit
$N\to\infty$ and analyze $Z[\vec{\psi}]$ by the saddle point
method. Since $N\to\infty$ and the stored patterns $\vec{\xi}^{\mu}$ are
random patterns, we assume self-averaging, such that $Z[\vec{\psi}]$ can
be replaced with its ensemble average $\overline{Z}[\vec{\psi}]$. We
can, therefore, obtain a rigorous solution by the path-integral method
\cite{Kawamura2002}.

Finally we obtain following macroscopic state equations from
$Z[\vec{\psi}]$ when $I_i(s)=0$:
\begin{eqnarray}
 m(s) &=& \left<\xi^{s}\int Dz F\left(\xi^{s}m(s-1)
           +z\sqrt{\alpha R(s-1,s-1)}\right)\right>_{\xi} ,
 \label{eqn:mt_xi}
\end{eqnarray}
\begin{eqnarray}
 R(s,s')&=& C(s,s') 
  +G(s,s-1)G(s',s'-1)R(s-1,s'-1) , 
 \label{eqn:R}
\end{eqnarray}
\begin{eqnarray}
 G(s,s-1) &=& \frac{1}{\sqrt{\alpha R(s-1,s-1)}}
  \left< \int \!Dzz F
   \left(\xi^{s}m(s-1)+z\sqrt{\alpha R(s-1,s-1)}\right)\! \right>_{\xi} ,
 \label{eqn:Gt_xi}
\end{eqnarray}
\begin{eqnarray}
 C(s,s')
 &=& \left<\int\frac{d\vec{z}}{2\pi\vert\vec{R}_{11}\vert^{\frac{1}{2}}}
      \exp\left[-\frac{1}{2}\vec{z}\cdot\vec{R}_{11}^{-1}\vec{z}\right]
	    F\left(\xi^{s}m(s-1)+\sqrt{\alpha}z(s-1)\right)
	  \right. \nonumber \\
 && \left. \times F\left(\xi^{s'}m(s'-1)+\sqrt{\alpha}z(s'-1)\right)
			 \right>_{\xi} ,
			 \label{eqn:Ct_xi} 
\end{eqnarray}
where $Dz=\frac{dz}{\sqrt{2\pi}}\exp[-\frac{1}{2}z^2]$, and
$\left<\cdot\right>_{\xi}$ denotes the average over all $\xi$'s. The
matrix $\vec{R}_{11}$ is a $2\times2$ matrix consisting of the elements
of $\vec{R}$ at time $s-1$ and time $s'-1$, and $\vec{z}=[z(s-1),
z(s'-1)]^T$. From equations (\ref{eqn:mt_xi})--(\ref{eqn:Ct_xi}),
$C(s,s')=0$ and $R(s,s')=0$ when $s\neq s'$.  Since $G(s,s-1)$ and
$C(s,s)$ can be described using only $m(s-1)$ and $R(s-1,s-1)$,
macroscopically this system is a two-degree-of-freedom system of $m(s)$
and $R(s,s)$.

Besides these dynamic macroscopic state equations, the fixed points of
the system are required in order to analyze the bifurcation of the
system. We set $m(t)\to m, G(t,t-1)\to G, R(t,t)\to r$ when
$t\to\infty$. Then, the previously obtained stationary state equations 
\cite{During1998,Katayama2001} are re-derived from our dynamic theory,
\begin{eqnarray}
 m &=& \left<\xi\int Dz F
        \left[\xi m+z\sqrt{\alpha r}\right] \right>_{\xi} ,
  \label{eqn:stat_m} \\
 G &=& \frac{1}{\sqrt{\alpha r}}\left<\int Dzz F
        \left[\xi m+z\sqrt{\alpha r}\right] \right>_{\xi} , \\
 r &=& \frac{1}{1-G^2} .
  \label{eqn:stat_r}
\end{eqnarray}

\begin{figure}[tb]
 \hfill
 (a) \includegraphics[width=65mm]{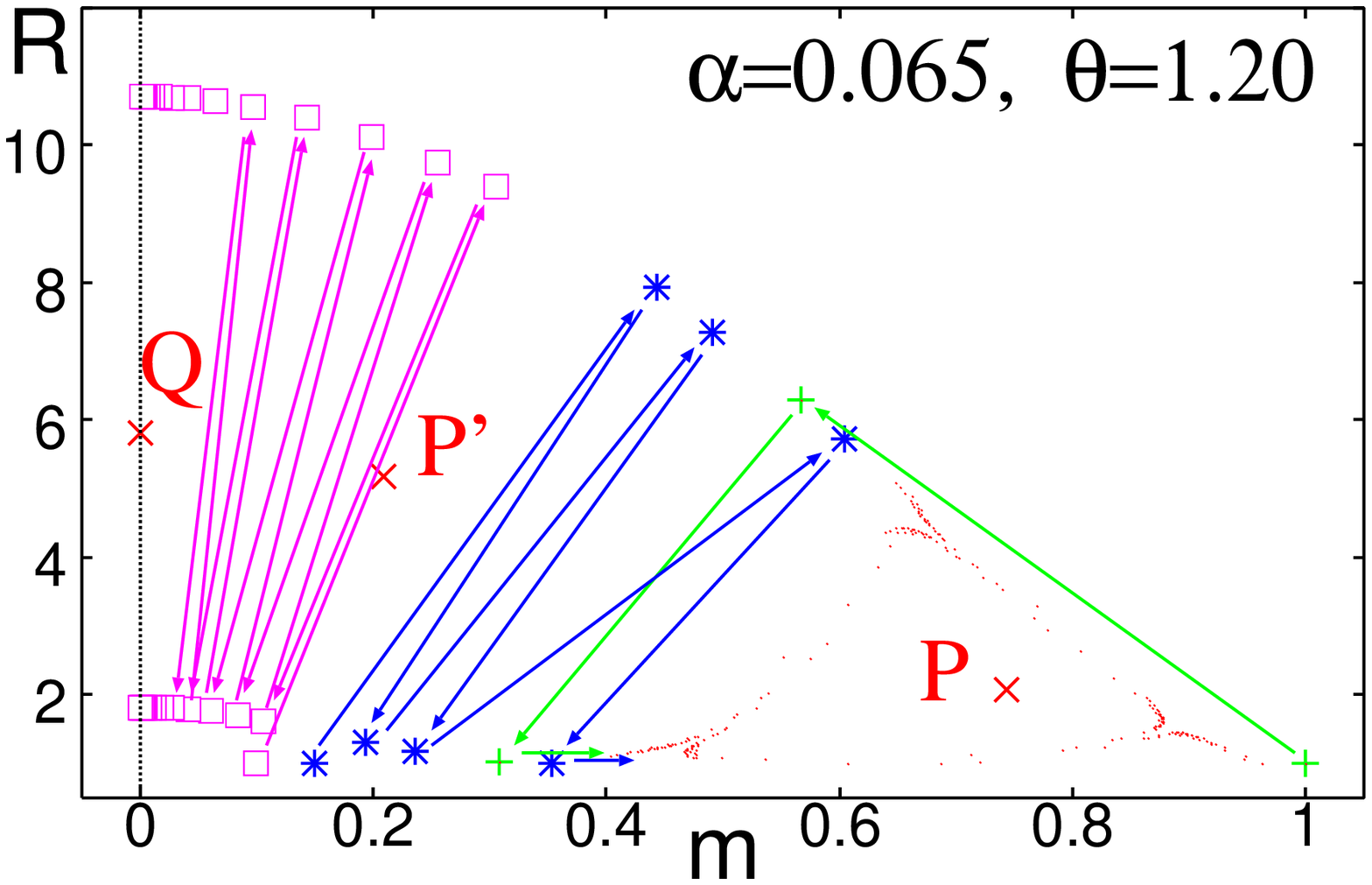}
 \hfill
 (b) \includegraphics[width=65mm]{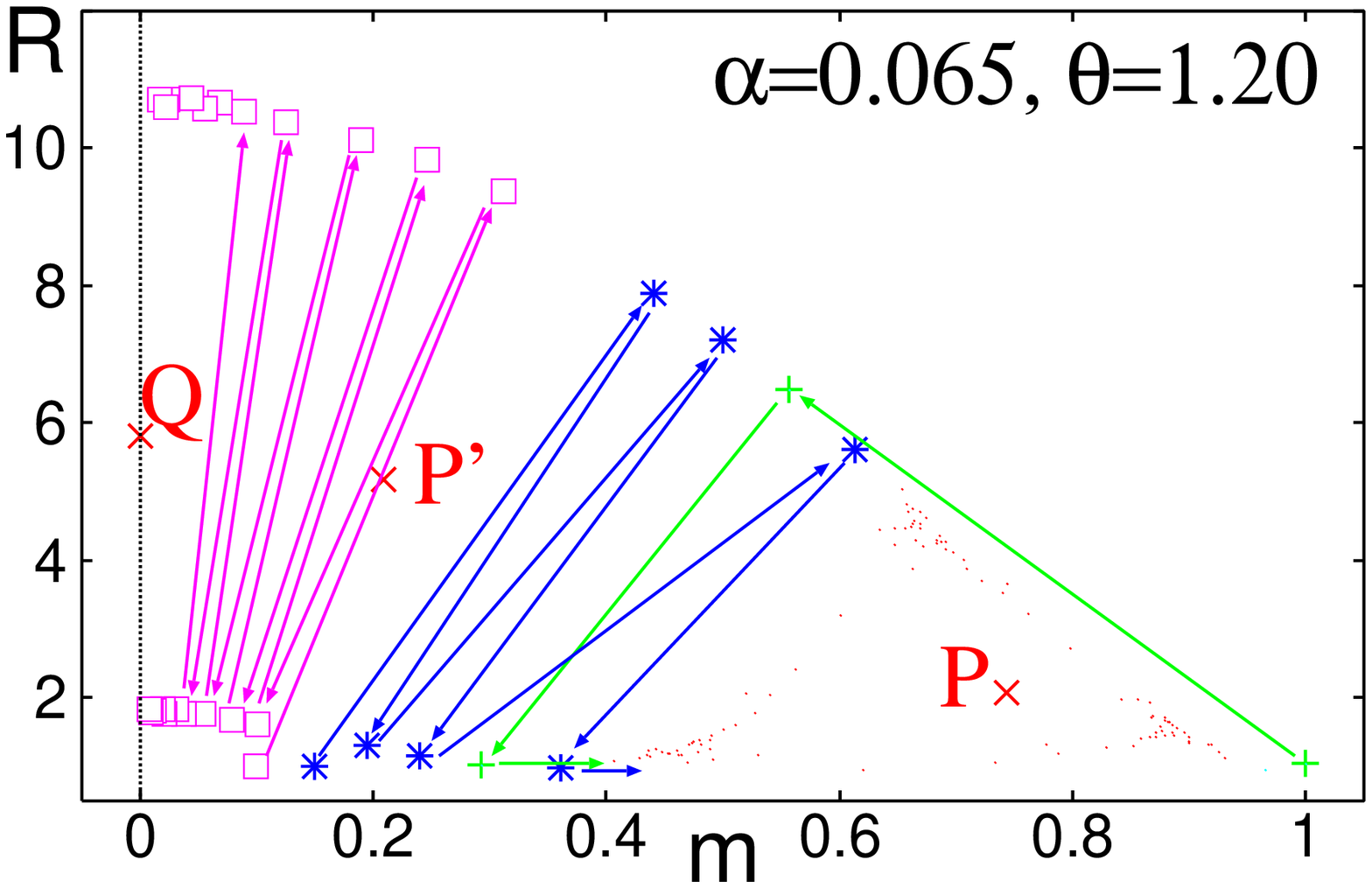}
 \hfill\mbox{}
 \caption{Transition of overlap $m(t)$ and variance of crosstalk noise
 $\alpha r(t)$ is shown for $\alpha=0.065, \theta=1.20, T=0$. 
 Fixed points $P,P',Q$ are from stationary state equations.
 (a) Results by theory. (b) Results by simulation.
 }
 \label{fig:mr}
\end{figure}

From our macroscopic state equations
(\ref{eqn:mt_xi})--(\ref{eqn:Ct_xi}), we analyze the transient dynamics
in the case of absolute zero temperature $T=0$.  Figure~\ref{fig:mr}
shows the transition of the overlap $m(t)$ and the variance of the
crosstalk noise $\alpha R(t,t)$. The figures show the results obtained
by (a) our theory, and (b) computer simulations with $N=100000$, where
the loading rate is $\alpha=0.065$ and the non-monotonicity
$\theta=1.20$. The cross marks $P,P',Q$ are fixed points.  The point $Q$
on the line $m=0$ is a saddle node (orientation reversing), $P'$ is a
repellor (unstable node), and $P$ is a repellor (unstable focus). There
is a period-$2$ attractor $Q_2$. The attractor $Q_2$ attracts the
trajectories with initial state $m(0)\approx 0$. Moreover, there is a
chaotic attractor around the repellor $P$. The Lyapunov exponent of the
chaotic attractor, which can be easily calculated from the equations
(\ref{eqn:mt_xi})--(\ref{eqn:Ct_xi}), is positive i.e. 0.038. The
results obtained by the theory agree with those of the computer
simulations.

\begin{figure}[bt]
 \begin{center}
   (a) \includegraphics[width=100mm]{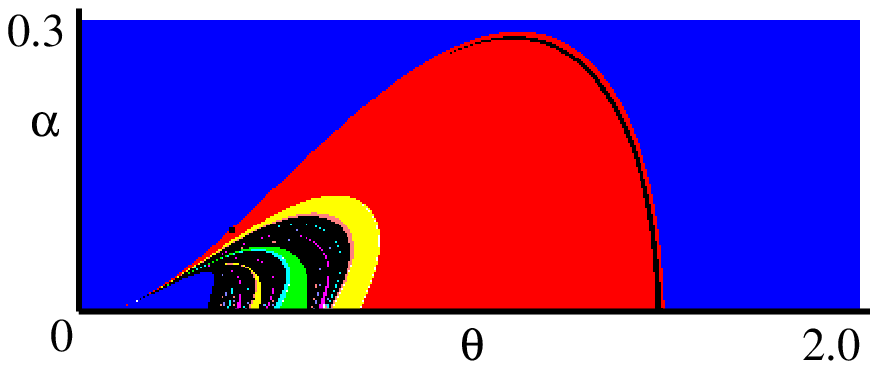}

   (b) \includegraphics[width=100mm]{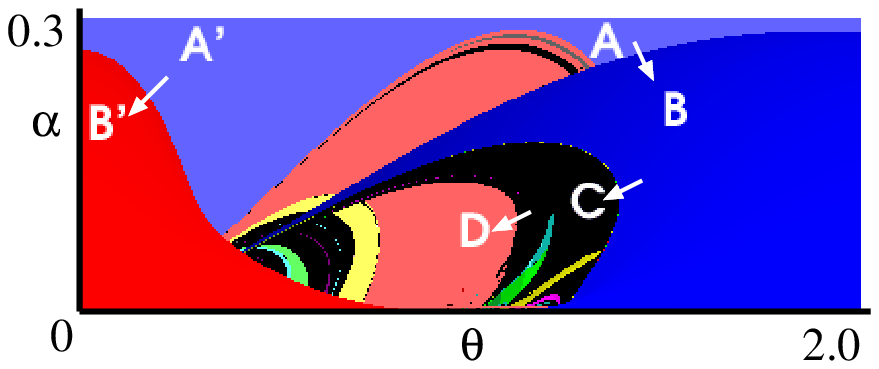}
 \end{center} 

 \caption{Two-parameter bifurcation diagram ($\theta,\alpha$) for (a)
 fixed point $Q$ and (b) fixed point $P$. The blue region stands for
 period-$1$ attractors, red for period-$2$, green for period-$3$, yellow
 for period-$4$, purple for period-$5$, sky blue for period-$6$, and
 black for more than six period attractors. } \label{fig:phase}
\end{figure}

The occurrence of invariant sets, as shown in Figure~\ref{fig:mr}, and
their relation are investigated in two-parameter space with respect
to $(\theta,\alpha)$. First, the line $m=0$ is a invariant set of the
macroscopic state equations, and the dynamic structure of this invariant
line obeys a one-dimensional map with respect to $R$ as
\begin{eqnarray}
 R(s,s)&=&1+
  \frac{2}{\pi\alpha}
  \left\{1-2\exp\left(-\frac{\theta^2}{2\alpha R(s-1,s-1)}\right)\right\}^2 .
  \label{eqn:R0}
\end{eqnarray}
Figure~\ref{fig:phase}(a) shows a two-parameter bifurcation diagram of
the invariant line $m=0$. This map satisfies the unimodal for almost all
regions in the parameter space $(\theta,\alpha)$. As $\theta$ decreases,
a period-$1$ attractor $Q$ bifurcates to a period-$2$ attractor $Q_2$
and evolves into a chaotic attractor by the period-doubling cascade. The
Lyapunov exponent at $\alpha=0.01$, $\theta=0.5$ is positive
i.e. 0.035. From the point of view of an associative memory, chaotic
behavior can be observed while the associative memory fails to retrieve
the stored patterns.

Next, we construct a two-parameter bifurcation diagram where the initial
state is away from the invariant line, i.e. $m\approx1$, and analyze the
coexistence with these attractors in the all-phase space.  In regions
$A$ and $A'$ in Figure~\ref{fig:phase}(b), there exists only a
period-$1$ attractor $Q$ on the invariant line. In region $D$, a
period-$1$ attractor $Q$ and a period-$2$ attractor $Q_2$ exist on the
invariant line. In these cases, since the stored patterns are all
unstable, the associative memory fails to retrieve from any initial
state.

In region $B$, both a period-$1$ attractor $Q$ on the invariant line and
a period-$1$ attractor $P$ near $m=1$ coexist. In this case, since the
stored patterns are stable, the associative memory can succeed in
retrieving when the state is in the basin of the attraction of $P$. Also
in region $B'$, both the attractor $Q$ and a period-$2$ attractor $P_2$
coexist.  The attractor $P_2$ is a sign-reversing state near the line
$m=\pm1$.  In this case, the stored patterns are unstable, and the
memory retrieves the stored pattern and its reverse one by turns when
the state is in the basin of attraction of $P_2$.

In region $C$, both a period-$2$ attractor $Q_2$ on the invariant line
and a quasi-periodic or chaotic attractor near $m=1$ coexist. In this
case, although the stored patterns are unstable, a quasi-periodic or
chaotic attractor exists. This state, therefore, goes to the attractor
instead of the memory state. Since the overlap is non-zero, the
associative memory neither completely succeeds nor fails to retrieve.

The coexistence, as stated above, can be explained by characteristic
bifurcations occurring on the boundary between the regions. On the
boundary $A\to B$, a saddle node $P'$ and a period-$1$ attractor $P$ are
generated via the saddle node bifurcation, and then both $Q$ and $P$
coexists. They are separated by the basin boundary constituted by
$P'$. On the boundary $A'\to B'$, similarly, a period-$2$ saddle node
$P_2'$ and a period-$2$ attractor $P_2$ are generated via the saddle
node bifurcation, and then both $Q$ and $P_2$ coexists. On the other
hand, on the boundary $B\to C$, a period-$1$ attractor $P$ evolves into
a repellor via the Hopf bifurcation, and a quasi-periodic attractor is
generated around the repellor. The quasi-periodic attractor is sometimes
phase-locked, and then it evolves into a more complex quasi-periodic
attractor via the Hopf bifurcation again. After that, the repellor
inside the quasi-periodic attractor evolves into a {\itshape snap-back
repellor} \cite{Marotte1978} and a belt-like chaos appears. Finally, the
belt-like chaos grows wider and becomes a thick chaotic attractor
including the repellor $P$. On the boundary $C\to D$, the chaotic
attractor vanishes by a {\itshape boundary crisis} \cite{Grebogi1982},
since it comes into contact with the basin boundary constituted by
$P'$. Therefore, in region $D$, only a period-$2$ attractor $Q_2$ on the
invariant line exists.

Let us discuss the mechanism of occurrence of chaos at the macroscopic
level in this model.  Only one order-parameter, $m(s)$, dominates the
macroscopic behaviors of the present system without the frustration of
interaction, that is, in the case that the number of the stored patterns
is finite ($\alpha=0$).  We can easily show that there is no chaotic
attractor in this case.  That is, the chaotic attractor appears when the
system has frustration, i.e., $\alpha\neq0$.  While the local field
$h_i(s)$ obeys distribution of $\delta$-function when $\alpha=0$, it
obeys the Gaussian distribution with the variance $\alpha R(s,s)$ when
$\alpha \neq 0$.  Because of this finite value of variance of the local
field distribution, the processing units, whose absolute values of the
local field $h_i(s-1)$ are around the non-monotonicity $\theta$, take
different values similarly to Bakers' Map.  This is intuitive reason
why the chaos occurs in this model with the frustration.  Non-trivial
things we found are that the chaos appears also in the case of
$\alpha\to0$ and the phase is completely different from the case of
$\alpha=0$.  Although the variance of the local field, $\alpha R(s,s)$,
converges to $0$ in the case of $\alpha\to0$, the order-parameter
$R(s,s)$ can take finite values and can behave chaotic in this case.

On the other hand, Sompolinsky et al.\cite{Sompolinsky1988} discussed
chaotic behaviors of a random neural network by the gain parameter $gJ$
with nonlinearity $g$ and the variance of couplings $J^2/N$. The
frustration exists in the network when $J>0$. For some $g$ the chaotic
behavior appears when $J>1/g$ and it does not appear when
$0<J<1/g$. That is, chaos dose not always appear even if the network has
frustration.  Therefore, we have found chaos in the frustrated system,
namely, {\itshape frustration-induced chaos}.

In summary, we considered the sequential associative memory model
consisting of non-monotonic units, which is a large-degree-of-freedom
system, and derived macroscopic state equations by the path-integral
method in the frustrated case. The results obtained by theory agreed
with the results obtained by computer simulations. We constructed
two-parameter bifurcation diagrams and explained the structure of the
bifurcations. The chaos at the macroscopic level can become
low-dimensional, since the macroscopic state equations obey the two
parameters. The chaos in the model is induced by the non-monotonicity
and the frustration. Since we would like to show that the occurrence of
the chaos is not caused by temperature but frustration, the case of
absolute zero temperature $T=0$ has been discussed in this Letter. The
case of finite temperature $T>0$ will be appeared elsewhere.









\end{document}